\documentclass[prl,twocolumn, superscriptaddress]{revtex4-1}%
\usepackage{amsfonts}
\usepackage{amsmath}
\usepackage{amssymb}
\usepackage{graphicx}%
\usepackage{soul}
\usepackage{color}
\providecommand{\U}[1]{\protect\rule{.1in}{.1in}}
\newcommand{\bra}[1]{\ensuremath{\left\langle#1\right|}}
\newcommand{\ket}[1]{\ensuremath{\left|#1\right\rangle}}

\begin{document}
\title{A Knob for Markovianity}

\author{Frederico Brito}
\email[]{fbb@ifsc.usp.br}
\affiliation{Instituto de F\'{i}sica de S\~ao Carlos, Universidade de S\~ao Paulo, C.P. 369, 13560-970, S\~ao Carlos, SP, Brasil}

\author{T. Werlang}
\affiliation{Instituto de F\'{\i}sica, Universidade Federal de Mato Grosso, 78060-900, Cuiab\'{a}, Mato Grosso, Brazil}

\begin{abstract}
We study the Markovianity of a composite system and its subsystems. We show how the dissipative nature of a subsystem's dynamics can be modified without having to change properties of the composite system environment. By preparing different system initial states or dynamically manipulating the subsystem coupling, we find that it is possible to induce a transition from Markov to non-Markov behavior, and {\it vice versa}.
\end{abstract}

\maketitle
\section{Introduction}
The theory of open quantum systems plays a central role in the description of realistic quantum systems due to unavoidable interaction with the environment. As is well known, the system-environment interaction can lead to energy dissipation and decoherence \cite{breuer}, posing a major challenge to the development of modern technologies based on quantum coherence \cite{nielsen}. Due to its fundamental character and practical implications, the investigation of dissipative processes has been a subject of vigorous research, where the standard approach assumes a system-environment weak coupling and a memoryless quantum dynamics (the Born-Markov approximation). Under such assumptions, system dynamics are determined by a quantum Markovian master equation, i.e., a completely positive quantum dynamical map with a generator in the Lindblad form \cite{breuer,lindblad}.

Although the Markovian approach has been widely used, there is a growing interest in understanding and controlling non-Markovianity. In quantum metrology, for example, entangled states can be used to overcome the shot noise limit \cite{Itano} in precision spectroscopy, even in the presence of decoherence \cite{Huelga95}. However, as suggested in Refs. \cite{chin12,yuichiro}, higher precision could be achieved in a non-Markovian environment, since a small Markovian noise  would be enough to restore the shot noise limit. Non-Markovian dynamics also play an important role in quantum biology \cite{fleming07}, where interaction with a non-Markovian environment can be used to optimize energy transport in photosynthetic complexes \cite{vega}, and can be observed in condensed matter devices like quantum dots \cite{reithmaier04,ubbelohde} and superconducting qubits \cite{yoshihara}. Furthermore, as pointed out recently in studies involving quantum key distribution \cite{vasile11}, quantum correlation generation \cite{huelga12}, optimal control \cite{schmidt11}, and quantum communication \cite{bylicka13}, the use of non-Markovian dynamics could offer an advantage over Markovian dynamics.  

This scenario has motivated studies aimed at characterizing and quantifying non-Markovian aspects of the time evolution of an open quantum system \cite{newworks}. However, unlike the classical case, the definition of non-Markovianity in the scope of quantum dynamics is still a controversial issue. For example, Breuer, Laine and Piilo (BLP) \cite{breuer09} have proposed a measure for non-Markovianity using the fact that all completely positive-trace preserving (CPTP) maps  increase the indistinguishability between quantum states. From a physical perspective, a quantum dynamics would be non-Markovian if there were a temporary back-flow of information from the environment to the system. On the other hand, for Rivas, Huelga and Plenio (RHP) \cite{plenio10}, a quantum dynamics would be non-Markovian if it could not be described by a {\it divisible} CPTP map. Formally, for such cases, one could not find a CPTP map $\Phi:\rho(0)\mapsto\rho(t)=\Phi(t,0)\rho(0)$, describing the evolution of the density operator $\rho$ from time $0$ to $t$, such that $\Phi(t+\tau,0)=\Phi(t+\tau,t)\Phi(t,0)$, where $\Phi(t+\tau,t)$ and $\Phi(t,0)$ are two CPTP maps. Therefore, the indivisibility of a map would be the signature of non-Markovian dynamics. These two different concepts of non-Markovianity are not equivalent \cite{mazzola10}: although all divisible maps are Markovian with respect to the BLP criterion, the converse is not always valid \cite{breuer12}.

In this paper, we explore the idea of how one might manipulate the Markovian nature of a dissipative subsystem, by exploiting features of its being a part of a composite system. For that, we study the dynamics of interacting two-state systems (TSS) coupled to a common thermal reservoir. By changing the composite initial state and/or the TSS couplings, we show that it is possible to modify {\it in situ} the characteristics of the subsystem's dissipation, enabling one to induce a transition from Markovian to non-Markovian dynamics and {\it vice versa}. Moreover, we observe the possibility of having different behaviors for the composite and subsystem, even when they are coupled to a common thermal environment.  Finally, we provide a qualitative and quantitative description of how the environmental TSS acts as part of the subsystem environment.

\section{The Model}
We initiate our analysis by choosing an exactly soluble analytical model that is capable of presenting the physics we want to exploit from dissipative composite systems. Therefore, our starting point is the dephasing model for two interacting two-state systems (2-TSS)
\begin{eqnarray}
\textstyle
H(t)=H_S(t)+\sum_k\hbar\omega_k b_k^\dagger b_k+\frac{S_z}{2}\sum_k\hbar (g^\ast_kb_k+g_kb_k^\dagger),~~~
\label{horiginal}
\end{eqnarray}
with $H_S(t)\equiv \sum_{i=1}^2\hbar\epsilon_i(t)\sigma_i^z/2+\hbar J(t)\sigma_1^z\sigma_2^z/2$, where $\sigma^z$ is the diagonal Pauli matrix and $S_z\equiv\sigma_1^z+\sigma_2^z$. The choice of this model is also motivated by the possibility of implementation in different experimental settings. For example, it could be realized in superconducting qubits \cite{harris}, trapped ions \cite{porras}, ultracold atoms in an optical lattice \cite{simmon11}, and NMR systems \cite{diogo}. In addition, such a model, without TSS-TSS couplings, is also considered as a paradigm of quantum registers \cite{reina}.

The bath of oscillators, introduced by the canonical bosonic creation and annihilation operators $b_k^\dagger$ and $b_k$, is characterized by its spectral density ${\cal J}(\omega)\equiv \sum_k|g_k|^2\delta(\omega-\omega_k)$\cite{CL}, and is responsible for imposing a nonunitary evolution for the 2-TSS. Since $[\sigma_i^z,H]=0$, the populations of the eigenstates of $(\sigma_1^z,\sigma_2^z)$ are constants of motion and the coupling with the environment solely induces random dephasing between any superposition of those eigenstates. The (2-TSS)-bath time evolution operator can be determined as
\begin{eqnarray}
\lefteqn{U(t,0)=e^{-i\left(\sum_i^2\frac{\sigma_i^z}{2}\int_0^{t}d\tau \epsilon_i(\tau)+\frac{\sigma_1^z\sigma_2^z}{2}\int_0^{t}d\tau J(\tau)+t\sum_k\omega_k b_k^\dagger b_k\right)}}\nonumber\\
&&\times  e^{itS_z^2\sum_k\frac{|g_k|^2}{4\omega_k^2}\left\{\omega_k-\frac{\sin\omega_k t}{t} \right\}} e^{\frac{S_z}{2}\sum_k\left\{G_k(t)b^\dagger_k-G_k^\ast(t) b_k\right\}},~~~~
\end{eqnarray}
with $G_k(t)\equiv g_k(1-e^{i\omega_k t})/\omega_k$. Consequently, if $\rho_{SB}(t)$ denotes the density matrix of the 2-TSS plus bath, then $\rho_{SB}(t)=U(t,0)\rho_{SB}(0)U^\dagger(t,0)$. Regarding the (2-TSS)-bath initial correlations, the initial state $\rho_{SB}(0)$ is hereafter assumed to be separable, i.e., $\rho_{SB}(0)=\rho_{S}(0)\otimes\rho_B$, where the bath is considered to be in equilibrium at temperature $T$ and therefore $\rho_B=e^{-\sum_k\hbar\omega_kb_k^\dagger b_k/k_BT}/Z$.

\subsection{2-TSS dynamics}
The dynamics of the open 2-TSS follows from the system's reduced density matrix, defined by $\rho(t)\equiv{\rm Tr}_B[\rho_{SB}(t)]$, where ${\rm Tr}_B[\dots]$ means the trace of bath degrees of freedom. Performing the calculation, one can find that $\rho(t)$ obeys the exact master equation
\begin{eqnarray}
&&\dot{\rho}(t)=\nonumber\\
&&~~-\frac{i}{\hbar}[H_S(t),\rho(t)]+\frac{\gamma(t)}{2}\left(S_z\rho(t) S_z-\frac{1}{2} \{S_z^2,\rho(t)\}\right),
\label{meq}~~
\end{eqnarray}
where $[~,~]$($\{~,~\}$) denotes the standard (anti)commutator, and $\gamma(t)\equiv\int d\omega\frac{{\cal J}(\omega)}{\omega}\sin\omega t\coth\frac{\hbar\omega}{2k_B T}$ is the time-dependent dephasing rate \footnote{Due to the system-environment interaction, the (unobservable) {\it bare} TSS-TSS $J$ coupling is renormalized, such that $J(t)\rightarrow J(t)+\int d\omega {\cal J}(\omega)\left(1-\cos\omega t \right)/\omega$, when working with the system's reduced dynamics. Throughout the manuscript, the $J$ values used for the system's reduced dynamics stand for their renormalized values.}. Since no approximation has been made whatsoever, it is worth noting that Eq. (\ref{meq}) constitutes a genuine CPTP quantum dynamical map.

The master equation (\ref{meq}) has a very suitable form for the analysis of quantum Markovianity, since it can be directly compared with the well-known Lindblad theory for open systems \cite{breuer,lindblad}. Indeed, if $\gamma(t)\geq0$ for all $t\geq0$, the time-local master equation describes a divisible CPTP map. Therefore, under this condition, the dynamics would fall into the class of problems considered as paradigms of quantum Markovian processes, since there is only a single decoherence channel \cite{breuer09,plenio10,breuer12,hall14}.    

\begin{figure}[]
\begin{center}\includegraphics[ width=1\columnwidth,
 keepaspectratio]{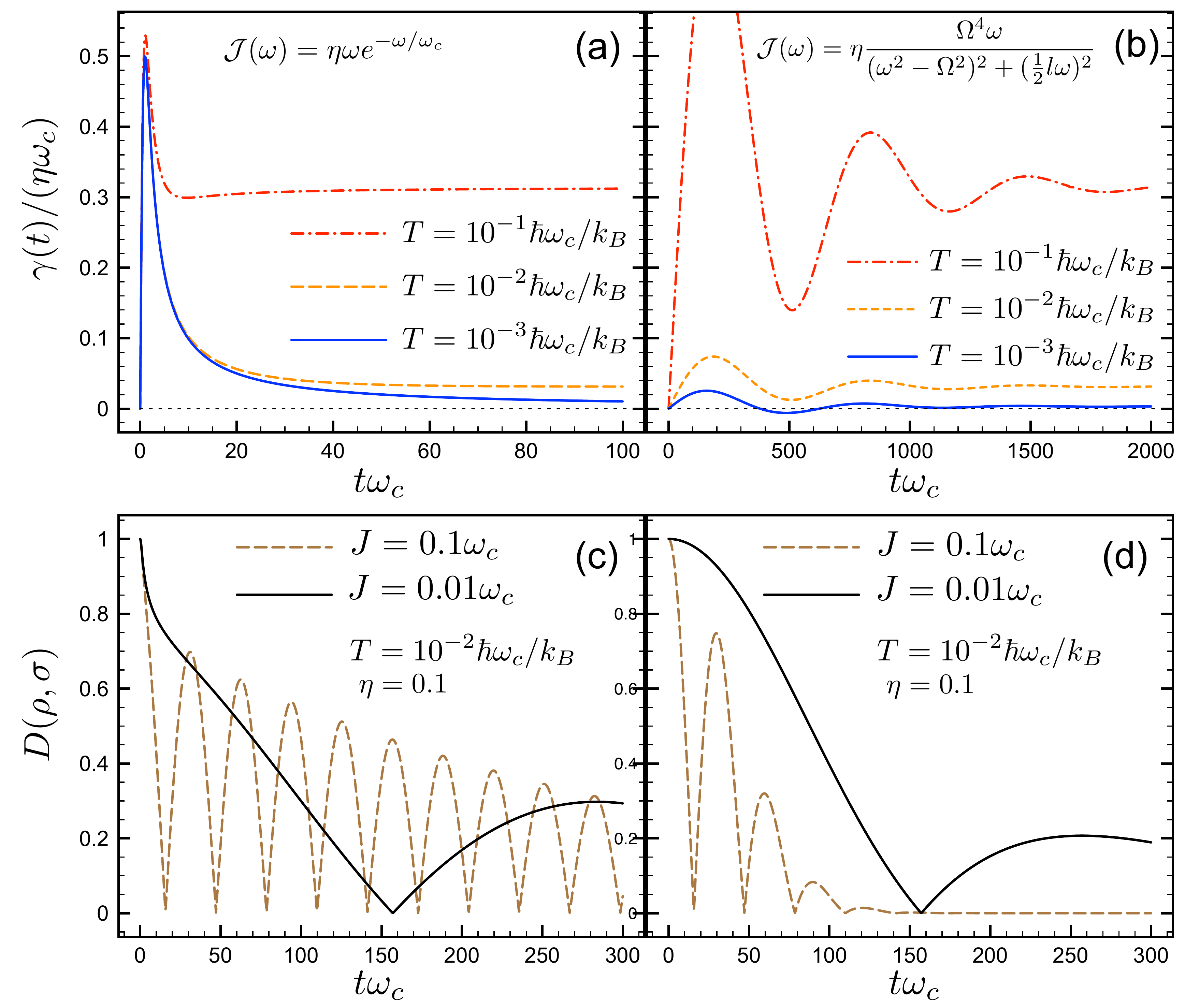}\end{center}
    \caption{The dephasing rate $\gamma(t)$ (a-b) and trace distance Eq. (\ref{tracered}) (c-d) as a function of time for the system Hamiltonian Eq. (\ref{horiginal}) in an Ohmic (a, c) and Lorentzian (b, d) environment. Depending upon the bath temperature $T$, it is found non-negative dephasing rates for all $t$ for both spectral densities, indicating regimes of Markov behavior for the 2-TSS dynamics. The trace distance is taken for the TSS reduced density matrix for two TSS-TSS constant coupling strengths $J$, considering a bath temperature $T$ where the 2-TSS dephasing rate has been found to be always positive, and orthogonal initial states given by $\ket{\Psi^a}=\frac{1}{2}(\ket{+}-\ket{-})\otimes (\ket{+}+\ket{-})$ and $\ket{\Psi^b}=\frac{1}{2}(\ket{+}+\ket{-})\otimes (\ket{+}+\ket{-})$. The non-monotonic behavior is a signature of non-Markov process for the single TSS dynamics. Physical parameters used for this plot: in addition to the values quoted in the panels, we set $\Omega=\l=0.01\omega_c$.}
    \label{fig}
\end{figure}

The simplest example one can find for the system Hamiltonian Eq. (\ref{horiginal}) that has $\gamma(t)\geq0, \forall t\geq0$, is the case of an ohmic bath, i.e., ${\cal J}(\omega)=\eta\omega e^{-\omega/\omega_c}$\footnote{For the class of parameterized spectral functions ${\cal J}(\omega)=\eta\omega^se^{-\omega/\omega_c}$, one finds that, if $s\leq2$, $\gamma(t)$ is a positive-valued function for any temperature and therefore the system dynamics is Markovian. On the other hand, if $s>2$, depending upon the temperature, $\gamma(t)$ can present negative values, given a non-divisible map.}. Indeed, as depicted in Fig. \ref{fig}(a), for any bath temperature $T$, the dephasing rate satisfies the condition of being non-negative for each fixed $t\geq0$. Consequently, the 2-TSS dissipative dynamics would be categorized as Markovian. 

Another important case happens when the environment seen by the system of interest presents a pronounced peak (often referred as a Lorentzian peak \cite{grifoni,bc}) at a characteristic frequency $\Omega$. Relevant examples are superconducting qubits coupled to readout dc-SQUIDS \cite{yoshihara}, electron transfer in biological and chemical systems \cite{bio} and semicondutor quantum dots \cite{qdots}, to name just a few. Figure \ref{fig}(b) shows the dephasing rate assuming a Lorentzian shape for the bath spectral density \cite{grifoni, bio}. As one can observe, for the case in which the resonance width $l$ is of the same order as the frequency peak $\Omega$, the dephasing rate can present negative values (solid curve) or be a positive-valued function (dashed and dot-dashed curves), depending upon the bath temperature. Thus the composite dissipative dynamics would be Markovian as long as the thermal energy scale is comparable to or larger than the resonance parameters (dashed and dot-dashed curves), i.e., $k_BT\gtrsim\hbar\Omega\sim\hbar l$.       

\subsection{Single TSS dynamics}
Thus far, we have only been concerned with characterizing the dissipative dynamics of the composite system, i.e., the open 2-TSS. What we now want to address is whether or not the single TSS dissipative dynamics can be tuned {\it in situ} such that its behavior would differ from that observed for the composite system. In fact, one could envision that, due to its interaction with the other TSS (hereafter labeled the {\it auxiliary} TSS), a single TSS would be coupled to a structured bath (auxiliary TSS+environment), which would play the role of an effective bath, and could induce a different nature for the dissipative process. If so, it might be possible to change dynamically the nature of such a structured bath by varying the TSS-TSS coupling $J$ and/or the initial state of the auxiliary TSS.       

As matter of fact, the approach of considering the subsystem's environment to be composed a common environment plus the rest of the composite system was employed long ago. Indeed, regarding to the characterization of an effective dissipative mechanism, it was successfully applied to the class of problems mapped onto a TSS coupled to a harmonic oscillator in the presence of a Markov bath, where perturbative \cite{bio, grifoni}, non-perturbative \cite{smith}, spectral density series representation \cite{hughes} and semi-infinite chain representation \cite{chin} techniques were proposed to describe such an effective dynamics in several contexts. In addition, a multi-spin environment coupled to local bosonic baths has also been studied \cite{lorenzo} in the context of Markovianity. It is also noteworthy that the presence of initial correlations in an environment composed of several subparts can lead to changes in the subsystem dynamics. Such an influence has been successfully investigated, with regard to the BLP measure, theoretically \cite{laine} and experimentally \cite{liu} for the specific case of non interacting TSS-TSS systems, which interact locally with correlated multimode fields. In spite of the results mentioned above, no systematic investigation of their Markovianity has been undertaken using both RHP and BLP measures. Nor have interacting TSS-TSS composite systems in the presence of thermal baths been analyzed, where tunability seems more natural for manipulating the dissipative dynamics nature of several physical implementations.       

We shall address this question by looking at the single TSS reduced density matrix $
\tilde{\rho}_1(t)\equiv{\rm Tr}_2{\rm Tr}_B[\rho_{SB}(t)]={\rm Tr}_2[\rho(t)]$,
where ${\rm Tr}_2[\dots]$ means the trace of the auxiliary TSS degrees of freedom. Its matrix representation in the $\sigma^z$ eigenbasis ($\sigma^z\ket{\pm}=\pm\ket{\pm}$) can be cast in the simple form\begin{eqnarray}
&&\tilde{\rho}_1(t)=\nonumber\\
&&~~\left(\begin{array}{cc}\alpha & e^{-i \int_0^td\tau \epsilon_1(\tau)}e^{-\Gamma(t)}\beta(t) \\e^{i \int_0^td\tau \epsilon_1(\tau)}e^{-\Gamma(t)}\beta^\ast(t) & 1-\alpha\end{array}\right),
\label{redmatrix}~~~
\end{eqnarray}
where $\Gamma(t)\equiv\int_0^tdt'\gamma(t')$ and the matrix elements are given by $\alpha\equiv\bra{++}\rho_s(0)\ket{++}+\bra{+-}\rho_s(0)\ket{+-}$ and $\beta(t)\equiv e^{i\int_0^td\tau J(\tau)}\bra{+-}\rho_s(0)\ket{--}+e^{-i\int_0^td\tau J(\tau)}\bra{++}\rho_s(0)\ket{-+}$. Hence, one finds that, in general, the dynamical map describing Eq. (\ref{redmatrix}) is non-linear, since it depends on the composite initial state. Indeed, such a feature becomes clear when one tries to write the reduced density matrix Eq. (\ref{redmatrix}) in a similar form to a Kraus representation \cite{nielsen}, i.e., $\tilde{\rho}_1(t)=\sum_{k=0,\pm}\Pi_k\tilde{\rho}_1(0)\Pi_k^\dagger$, with operation elements $\Pi_k$ given by
\begin{widetext}
\begin{eqnarray*}
\Pi_0&=&e^{-\frac{i}{2} \int_0^td\tau \epsilon_1(\tau)}e^{-\frac{\Gamma(t)}{2}}\sqrt{\frac{\beta(t)}{\beta(0)}}\ket{+}\bra{+}+e^{\frac{i}{2} \int_0^td\tau \epsilon_1(\tau)}e^{-\frac{\Gamma(t)}{2}}\sqrt{\left(\frac{\beta(t)}{\beta(0)}\right)^\ast}\ket{-}\bra{-},\\
\Pi_\pm&=&\sqrt{1-e^{-\Gamma(t)}\left|\frac{\beta(t)}{\beta(0)}\right|}\ket{\pm}\bra{ \pm},~~{\rm with}~~ \sum_k\Pi_k^\dagger\Pi_k=\mathbf{1}.
\end{eqnarray*}
\end{widetext}Observe that those operators are dependent on the composite initial state, which would not be a genuine Kraus representation. Therefore, it is clear that, in general, the dynamical map describing the evolution of the single TSS reduced matrix $\tilde{\rho}_1(t)$ is itself dependent on the initial condition $\tilde{\rho}_1(0)$, which {\it breaks} the linearity of the map. Furthermore, the reduced dynamics are not necessarily described by a completely positive dynamical map. It is worthy of mentioning that, under such circumstances, the characterization of the single TSS dynamics as Markovian or not is disputable with current measures, since both RHP and BLP proposals rely on linear maps. Nevertheless, it is still possible to find conditions where the dynamical map of the single TSS reduced matrix $\tilde{\rho}_1(t)$ is a linear CPTP map, thus being consistent with RHP and BLP constructions. Those happen when i) the composite initial state is a product state, i.e., $\rho_s(0)=\rho_1(0)\otimes\rho_2(0)$, and ii) there is no TSS-TSS interaction ($J=0$), since for both cases one finds $\beta(t)/\beta(0)=1$. In other words, in these cases the reduced dynamics is described by a genuine Kraus representation.

It follows from Eq. (\ref{redmatrix}) that the master equation for the single TSS reduced density matrix reads
\begin{eqnarray}
\textstyle
\dot{\tilde{\rho}}_1(t)=-\frac{i}{\hbar} [\tilde{H}_1(t),\tilde{\rho}_1(t)]+\frac{\tilde{\gamma}(t)}{2}\Big(\sigma_1^z\tilde{\rho}_1(t) \sigma_1^z-\tilde{\rho}_1(t)\Big),~~
\label{redmeq}
\end{eqnarray}
where $\tilde{H}_1(t)\equiv\hbar(\epsilon_1(t)+\tilde{J}(t))\sigma^z_1/2$, with $\tilde{J}(t)\equiv\Im\left(\frac{\beta}{|\beta|^2}\frac{d\beta^\ast}{dt}\right)$, and $\tilde{\gamma}(t)\equiv\gamma(t)-\Re\left(\frac{\beta}{|\beta|^2}\frac{d\beta^\ast}{dt}\right)$. Note the Lindblad structure of the master equation (\ref{redmeq}), where $\tilde{H}_1(t)$ and $\tilde{\gamma}(t)$ play the roles of effective single TSS Hamiltonian and dephasing rate, respectively, and are manifestly dependent on the composite system initial state condition. Moreover, since $\beta$ depends only on the initial state $\rho_s(0)$ and the TSS-TSS coupling $J$, neither the unitary part of Eq. (\ref{redmeq}) nor the extra term in its dephasing rate is influenced by the system-bath coupling. Consequently, the bath and auxiliary TSS contributions for each term of Eq. (\ref{redmeq}) can be identified and quantified. 

Two results immediately become apparent: i) as expected, if the TSS-TSS coupling is zero ($J=0\rightarrow d\beta(t)/dt=0\rightarrow\tilde{J}=0~{\rm and}~\tilde{\gamma}=\gamma$), the single TSS dissipative dynamics is completely enforced by the environment;  and ii) if the auxiliary TSS initial state is set in an eigenstate of $\sigma^z$, one finds that $\Re\left(\frac{\beta}{|\beta|^2}\frac{d\beta^\ast}{dt}\right)=0\rightarrow\tilde{\gamma}=\gamma$ and $\tilde{J}=J\langle\sigma_2^z\rangle$, where $\langle\sigma_2^z\rangle\equiv {\rm Tr}[\rho_2(0)\sigma_2^z]$. These results highlight the fact that the auxiliary TSS dynamics will be locked into its initial state, since $\sigma^z$ is a constant of motion. Therefore the interacting term $J\sigma_1^z\sigma_2^z$ will only play the role of a fixed external field for the single TSS, in which dissipative dynamics would follow the process imposed by the environment.  Thus, for such cases, both the composite and single TSS dissipative dynamics would have the same behavior. 

In spite of the two cases discussed above, in general $\tilde{\gamma}$ is not determined only by the environment dephasing rate $\gamma$ and is not a positive-valued function. Therefore, with regard to the concordance between BLP and RHP measures, it is not possible to infer from the master equations (\ref{meq}) and (\ref{redmeq}) whether the single TSS dissipative dynamics will follow the same behavior as that observed for the composite system. Consequently, the two measures have to be determined in order to find the cases of agreement in our system. 

As already mentioned, the figure of merit for BLP is the distinguishability of quantum states \cite{breuer}. An important tool that has been used as a measure for the distinguishability of quantum states is the trace distance $D(\rho,\sigma)\equiv\frac{1}{2}{\rm Tr}|\rho-\sigma|$ between two quantum states $\rho$ and $\sigma$ \cite{nielsen}. According to BLP, since the trace distance has the feature that under CPTP maps $\Phi(t,0)$ its value cannot increase beyond its initial value \cite{nielsen}, i.e., $D(\rho_a(t),\rho_b(t))\leq D(\rho_a(0),\rho_b(0))$, where $\rho_{a,b}(t)=\Phi(t,0)\rho_{a,b}(0)$, the trace distance could also be used as a definition of non-Markovianity \cite{breuer09,breuer12}. That definition is based upon the idea that a Markovian dynamics has to be a process in which any two quantum states become less and less distinguishable as the dynamics flows, leading necessarily to a monotonic decrease of its trace distance. Thus, a non-monotonic behavior is interpreted as a back-flow of information from the environment to the system. From an experimental perspective, the trace distance could be calculated by the tomography of the density operator, as was recently done by Bi-Heng Liu {\it et al.} \cite{liu, liu11} using all-optical experimental setups, or inferred from current fluctuations \cite{ubbelohde}.

For states given by Eq. (\ref{redmatrix}), the trace distance can be easily determined as
\begin{eqnarray}
D(\tilde{\rho}_1^a(t),&&\tilde{\rho}_1^b(t))=\nonumber\\
&&\sqrt{(\alpha_a-\alpha_b)^2+e^{-2\Gamma(t)}|\beta_a(t)-\beta_b(t)|^2},
\label{tracered}~~~~~~
\end{eqnarray} 
which time behavior is manifestly dependent on the 2-TSS initial state. For instance, panels (c) (Ohmic) and (d) (Lorentzian bath) of Fig. \ref{fig} show a representative initial condition, in which the auxiliary TSS is set in an equal superposition of $\sigma^z$ eigenstates, i.e., $(\ket{+}+\ket{-})/\sqrt{2}$. The bath temperature is chosen such that both Ohmic and Lorentzian baths produce a Markovian process for the composite system. However, even though it asymptotically vanishes for both cases, the trace distance for single TSS states is non-monotonic, indicating a non-Markovian process. Moreover, this non-monotonic behavior implies that the TSS dynamics is indivisible \cite{breuer12}. Thus the TSS dynamics is non-Markovian not only in terms of the back-flow of information, but also with respect to the RHP criterion.   

\begin{figure}[]
\begin{center}\includegraphics[ width=1\columnwidth,
 keepaspectratio]{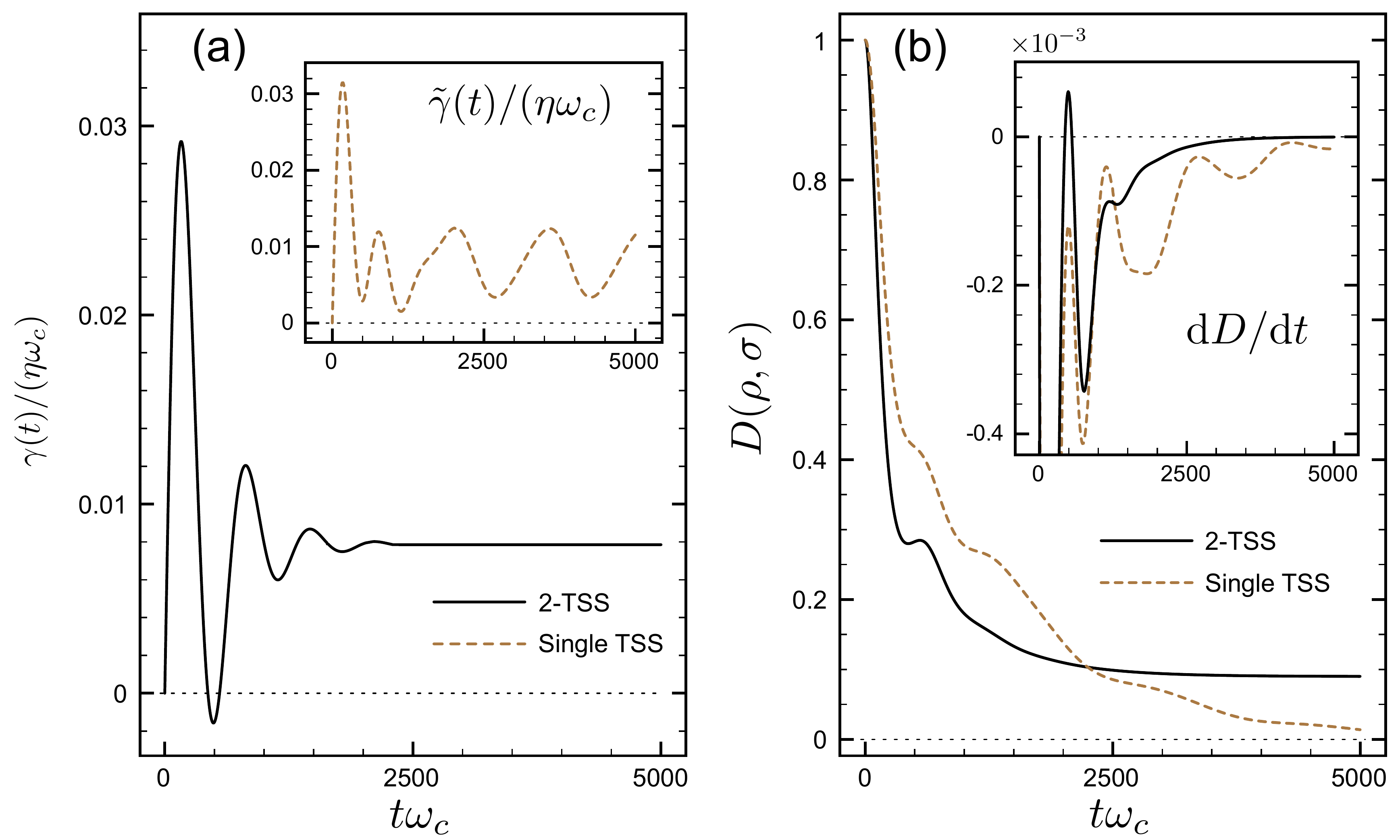}\end{center}
    \caption{{\it Panel a}: The dephasing rate for 2-TSS and a single TSS (inset) as a function of time for the system Hamiltonian Eq. (\ref{horiginal}) in a Lorentzian environment. Although the dephasing rate $\gamma(t)$ for 2-TSS assumes negative values, the effective dephasing rate $\tilde{\gamma}(t)$ for the single TSS is always positive. To compute $\tilde{\gamma}(t)$ we set the auxiliary TSS state as $\ket{\psi_{aux}}= \sqrt{0.1}\ket{+}+\sqrt{0.9}\ket{-}$. {\it Panel b}: Trace distance and its derivative (inset) for 2-TSS and a single TSS as a function of time for 2-TSS orthogonal initial states $\ket{\Psi^a}=\frac{1}{\sqrt{2}}(\ket{+}-\ket{-})\otimes \ket{\psi_{aux}}$ and $\ket{\Psi^b}=\frac{1}{\sqrt{2}}(\ket{+}+\ket{-})\otimes \ket{\psi_{aux}}$. The non-monotonic behavior of the trace distance is observed only for 2-TSS dynamics. Physical parameters used for this plot: $J=2\times 10^{-3}\omega_c$, $T=2.5\times10^{-3}\hbar\omega_c/k_B$, $\eta=0.1$ and $\Omega=\l=0.01\omega_c$.}
    \label{fig2}
\end{figure}

On the other hand, one can find a situation in which the single TSS dynamics is Markovian even when the composite system presents a non-Markovian behavior. This unexpected scenario is shown in Fig. \ref{fig2}. The negative values assumed by the dephasing rate $\gamma(t)$, panel (a), ensure that the 2-TSS dynamics is indivisible. Furthermore, as shown in panel (b), the trace distance can be found having a non-monotonic behavior. Therefore, the composite system conforms to a non-Markovian quantum process with respect to the BLP and RHP criteria. However, since the effective rate $\tilde{\gamma}(t)$ is always positive for the fixed initial condition $\rho_2(0)$, inset of panel (a), we have a divisible dynamics, implying  that the trace distance is a monotonic function of time \cite{breuer12}, as illustrated in panel (b). Consequently, the single TSS dynamics is a Markovian process in the standard way. A similar result was obtained for non-interacting TSS-TSS systems coupled locally with correlated multimode fields \cite{laine,liu}.

The results and examples offered here demonstrate i) that the single TSS hamiltonian has no influence regarding the Markovianity of both single TSS and composite system; ii) if one tries to push the characterization of the Markovianity of the single TSS dynamics for cases where the map is non-linear, it is found that a 2-TSS entangled state is not a sufficient condition for non-Markovianity of the single TSS dynamics. Indeed, a simple example would be having one of the Bell states, e.g., $\ket{\Phi_+}=(\ket{++}+\ket{--})/\sqrt{2}$, as the 2-TSS initial state. For those states, one finds $\beta(t)=0$ in Eq. (\ref{redmatrix}), which leads to a non-dissipative dynamics for the single TSS dynamics. This result explicitly shows that the single TSS reduced dynamics has a convoluted dependence with the 2-TSS initial state, which does not allow one drawing immediate conclusions about the role played by the presence of entanglement in the initial state; and iii) that the presence of the auxiliary TSS not only has quantitative effects, but  may also have, with regard to BLP and RHP criteria, a decisive impact on the nature of the TSS dynamics. Such an influence can be made more explicit if one focuses on cases where the composite initial state is a product state ($\rho_s(0)=\rho_1(0)\otimes\rho_2(0)$). In fact, for those the extra term in $\tilde{\gamma}$, namely, $\gamma_{aux}(t)\equiv-\Re\left(\frac{\beta}{|\beta|^2}\frac{d\beta^\ast}{dt}\right)$,  can be written in the simple form
\begin{eqnarray}
\gamma_{aux}(t)=\frac{J(t)}{2} \frac{\left(1-\langle\sigma_2^z\rangle^2\right)\sin\left(2\int_0^td\tau J(\tau)\right)}{1-\left(1-\langle\sigma_2^z\rangle^2\right)\sin^2\left(\int_0^td\tau J(\tau)\right)}.
\label{gammaaux}
\end{eqnarray}
Furthermore, if one considers problems having the same $\rho_2(0)$, the figure of merit for both criteria becomes $\tilde{\gamma}$, since the monotonicity of the trace distance Eq. (\ref{tracered}) is determined by $d D/dt\propto -\tilde{\gamma}$. For this case, the agreement between the two criteria is guaranteed because Eq. (\ref{redmeq}) describes a single decoherence channel with the same $\tilde{\gamma}(t)$ for all $\rho_1(0)$ \cite{hall14}. Thus, as $\tilde{\gamma}\equiv\gamma+\gamma_{aux}$, it is clear that can be established a competition when setting the nature of the reduced TSS dynamics, which can be tailored through the knobs $J$ and $\langle\sigma_2^z\rangle$ due to the presence of the auxiliary TSS.    

Indeed, except for the trivial case $|\langle\sigma_2^z\rangle|=1$, Eq. (\ref{gammaaux}) shows that the term due to coupling with the auxiliary TSS will induce an {\it ad infinitum}  pattern of momentary loss and recurrence of quantum coherence observed for the reduced TSS dynamics, which is a characteristic of the entanglement created between a tiny number of degrees of freedom. Thus the coupling with the auxiliary TSS constitutes a channel, here reversible, for exchanging information. The reversibility of such a channel is only possible here because of the diagonal nature of the interactions, which maintains the amplitude of $\gamma_{aux}$ constant. Consequently, the irreversibility of the dephasing process observed for the reduced TSS is only led by its direct coupling to the environment. Following that reasoning, one could expect that the same results would be found for the reduced TSS dynamics if an independent bath model were assumed, instead of the common environment model Eq. (\ref{horiginal}). As a matter of fact, as long as the diagonal nature of the interactions is preserved, the same results Eqs. (\ref{redmatrix})-(\ref{gammaaux}) are obtained, with $\gamma(t)$ reading the dephasing rates due to the individual baths coupled to the single TSSs.

A final note on the single TSS dissipative dynamics comes from the observation that, if $\rho_2(0)$ is assumed to be a thermal state, our problem resembles very much with the one having a unique (structured) thermal bath initially decoupled from the single TSS. Despite several known examples of the TSS-harmonic oscillator problem \cite{grifoni,bc,bio, qdots,smith,hughes,chin}, because of the different character between the spin and bosonic degrees of freedom, it does not seem possible to assign an effective spectral density for the TSS-TSS case. Nevertheless, for the case of a constant $J$ TSS-TSS interaction, one can single out a spectral density due to the spin character of the structured bath. Following from Eq. (\ref{gammaaux}), one finds that $\gamma_{aux}(t)=\int d\omega\frac{{\cal J}_{aux}(\omega)}{\omega}\sin\omega t\left(\frac{1-|\langle\sigma_2^z\rangle|}{1+|\langle\sigma_2^z\rangle|}\right)^{\omega/2J}=\int d\omega\frac{{\cal J}_{aux}(\omega)}{\omega}\sin\omega t\exp\left(-\frac{\epsilon_2}{J}\frac{\hbar\omega}{2k_BT}\right)$, which leads to the spin component of the spectral density
${\cal J}_{aux}(\omega)\equiv(2J)^2\sum_{m=1}^\infty(-1)^{m+1}m\delta(\omega-2Jm)$. Those finds reveal that the auxiliary TSS behaves as a mode filter for the natural $2J$ frequency and its harmonics, which are weighted by a Boltzmann factor of effective temperature $T_{spin}\equiv(2J/\epsilon_2)T$.

\section{Non-commuting system-environment interaction: A case study}
The analysis of the exact soluble model Eq. (\ref{horiginal}) gives evidence that the presence of an auxiliary system coupled to the system of interest could create a knob to manipulate {\it in situ} its Markovianity. However, since it was obtained for the particular situation where the system-environment interaction commutes with the system Hamiltonian, one could wonder whether such a feature would be spoiled if the system-bath interaction would not commute with the system Hamiltonian. Unfortunately, for such cases, exact solutions are rare, if not impossible. Nevertheless, we now present a case study, which is representative of some physical systems \cite{porras,simmon11}, where one finds that the idea of having a knob for Markovianity, due to the presence of an auxiliary system, does also apply.

Our case study has the feature that, {\it by construction}, the composite system dynamics is assumed to be Markovian, but depending on its initial state, the single TSS dynamics can be found to be both Markovian or non-Markovian. The Hamiltonian of interest is exactly the one given by Eq. (\ref{horiginal}), except for the system-environment interaction, which now reads
\begin{eqnarray}
\textstyle
H_{int}=\frac{S_x}{2}\sum_k\hbar (g^\ast_kb_k+g_kb_k^\dagger).~~~
\label{hmodified}
\end{eqnarray}
Assuming that the system-environment interaction is weak and the composite system dynamics is Markovian, such that the Born-Markov approximation is aplicable, the system's reduced master equation is found to have a Lindblad form with positive rates \cite{breuer}.

\begin{figure}[]
\begin{center}\includegraphics[ width=1\columnwidth,
 keepaspectratio]{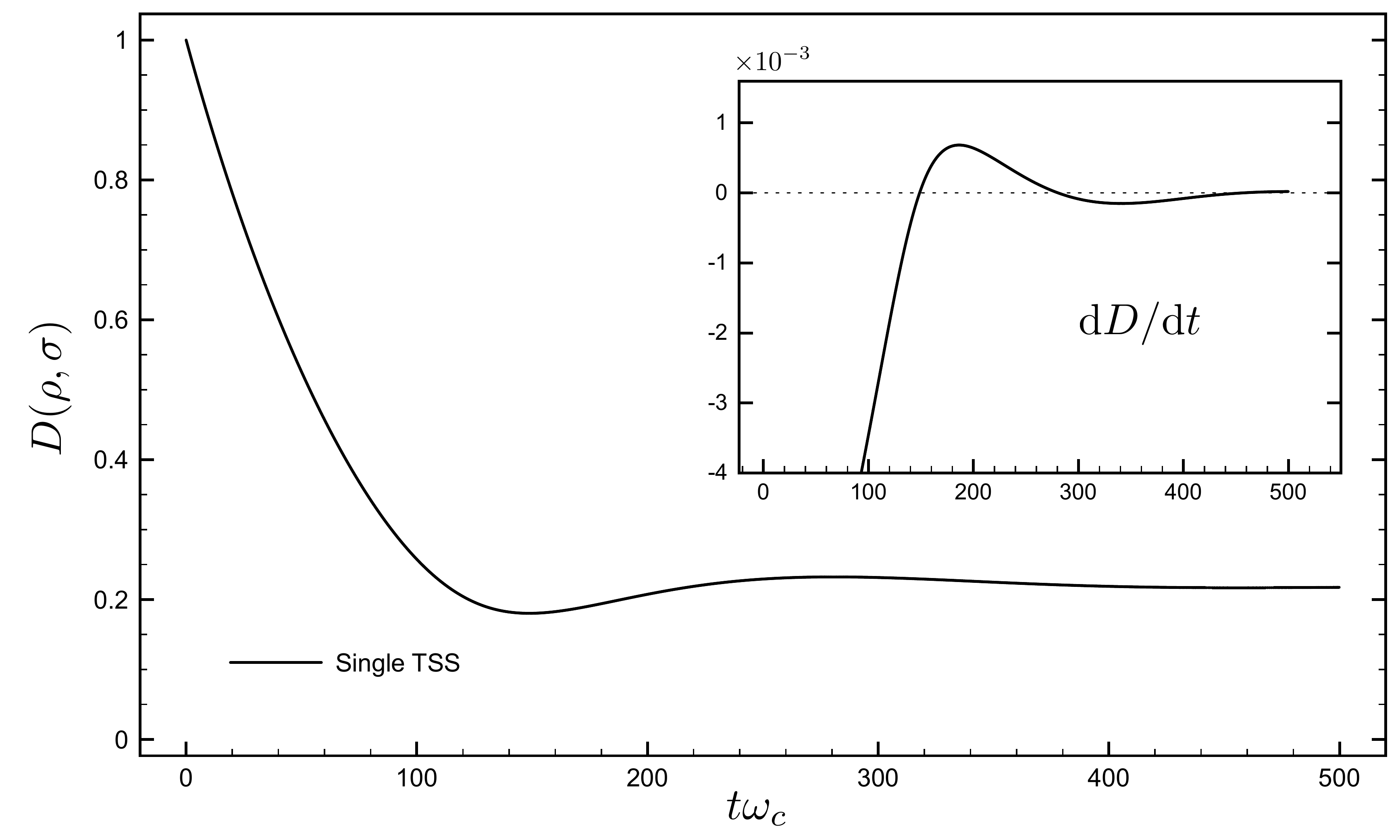}\end{center}
    \caption{Dynamics of the trace distance and its derivative (inset) for a single TSS calculated for an example (Eq. \ref{hmodified}) of a non-commuting system-environment interaction. Its non-monotonicity reveals that the single TSS dynamics is non-Markovian, whereas the 2-TSS dynamics is found to be Markovian (see main text). Here, it was assumed an Ohmic bath at temperature of $T=0{\rm K}$, and the 2-TSS orthogonal initial states $\ket{\Psi^a}=\frac{1}{\sqrt{2}}(\ket{+}-\ket{-})\otimes \ket{\psi_{aux}}$ and $\ket{\Psi^b}=\frac{1}{\sqrt{2}}(\ket{+}+\ket{-})\otimes \ket{\psi_{aux}}$, where we set the auxiliary TSS state as $\ket{\psi_{aux}}= 0.9\ket{+}+\sqrt{0.19}\ket{-}$. Physical parameters used for this plot: $J=0.01\omega_c$, $\eta=0.1$ and $\epsilon_1=\epsilon_2=0.1\omega_c$.}
    \label{fig3}
\end{figure}

As for the single TSS dynamics, as expected, its Markovian character is dependent on the 2-TSS initial state. For example, for the environment temperature $T=0{\rm K}$, if the 2-TSS initial state is the separable state $\rho_1(0)\otimes\ket{-}\bra{-}$, one can put the single TSS master equation in a Lindblad form given by
\begin{widetext}
\begin{eqnarray}
\tilde{\rho}_1(t)=-i\left(\epsilon-J\right)[\sigma_+\sigma_-,\tilde{\rho}_1] + \left(\frac{\gamma_-e^{-\gamma_- t}}{1+e^{-\gamma_- t}}\right)\left(2\sigma_-\tilde{\rho}_1\sigma_+ -\tilde{\rho}_1\sigma_+\sigma_--\sigma_+\sigma_-\tilde{\rho}_1\right),
\end{eqnarray}
\end{widetext}
where $\gamma_-(t)={\cal J}(\epsilon-J)>0$ \footnote{Here, for simplicity, we have assumed $\epsilon_1(t)=\epsilon_2(t)=\epsilon$, $J(t)=J$ and $\epsilon>J$. It is worthy of note that, for a general initial state, even for a separable state, it may not be possible writing the single TSS master equation in a Lindblad form.}. Thus, for that initial condition, the single TSS dynamics would be Markovian. On other hand, if the 2-TSS initial state is still separable, but given by another condition, e.g., $\rho_1(0)\otimes \ket{\psi_{aux}}\bra{\psi_{aux}}$, with $\ket{\psi_{aux}}= 0.9\ket{+}+\sqrt{0.19}\ket{-}$, one finds, using the trace distance, that the single TSS dynamics would be characterized as non-Markovian, according to both BLP and RHP criteria (see Fig. \ref{fig3}). Therefore, such examples make clear that the 2-TSS initial state can be used as a knob to determine the Markovianity of the single TSS, even having the system-environment interaction not commuting with the system dynamics.

\section{Conclusions and perspectives}
In summary, we have shown the possibility of manipulating the Markovianity of a subsystem of interest without having to change the common thermal environment properties or to assume a correlated environment state, which can be daunting requirements for some physical implementations. By choosing an exactly soluble model and a case study for a non-commuting system-environment interaction, we illustrated how one can induce a transition from Markovian to non-Markovian dynamics, and {\it vice versa}, by changing the characteristics of the composite system. Such analises build evidences that a knob for Markovianity can be introduced when one couples an auxiliary system to the system of interest. Among the perspectives offered by this work, one could envision points regarding  whether it would be possible having the concept of Markovianity and non-Markovianity defined for those cases which dynamics is described by non-linear maps, the extension to other non-commutative TSS-TSS interactions and the role of the number of degree of freedom involved in the coupling auxiliary system. On this last point, it would be interesting to investigate whether there would be a limitation on the maximum number of degrees of freedom allowing for the existence of the knob\\

\begin{acknowledgements}
F.B. is supported by Instituto Nacional de Ci\^encia e Tecnologia - Informa\c{c}\~ao Qu\^antica (INCT-IQ) and by Funda\c{c}\~ao de Amparo \`a Pesquisa do Estado de S\~ao Paulo  (FAPESP) under grant number 2012/51589-1. T.W. acknowledges financial support from CNPq (Brazil) through Grant No. 478682/2013-1.
\end{acknowledgements}

\end{document}